\documentclass[12pt]{iopart}
\usepackage{graphicx}
\usepackage{subfigure}
\usepackage{hyperref}
\hypersetup{colorlinks=true,breaklinks,urlcolor=blue,linkcolor=blue,citecolor=blue}

\usepackage{iopams}
\begin{document}

\title[Real spectra, Anderson localization, and topological phases in one-dimensional quasireciprocal systems]{Real spectra, Anderson localization, and topological phases in one-dimensional quasireciprocal systems}

\author{Qi-Bo Zeng$^1$, Rong L\"u$^{2,3}$}

\address{$^1$ Department of Physics, Capital Normal University, Beijing 100048, China}
\address{$^2$ State Key Laboratory of Low Dimensional Quantum Physics, Department of Physics, Tsinghua University, Beijing 100084, China}
\address{$^3$ Frontier Science Center for Quantum Information, Beijing, China}
\ead{zengqibo@cnu.edu.cn}

\vspace{10pt}
\begin{indented}
\item[]
\end{indented}

\begin{abstract}
We introduce the one-dimensional quasireciprocal lattices where the forward hopping amplitudes between nearest neighboring sites $\{ t+t_{jR} \}$ are chosen to be a random permutation of the backward hopping $\{ t+t_{jL} \}$ or vice versa. The values of $\{ t_{jL} \}$ (or $\{t_{jR} \}$) can be periodic, quasiperiodic, or randomly distributed. We show that the Hamiltonian matrices are pseudo-Hermitian and the energy spectra are real as long as $\{ t_{jL} \}$ (or $\{t_{jR} \}$) are smaller than the threshold value. While the non-Hermitian skin effect is always absent in the eigenstates due to the global cancellation of local nonreciprocity, the competition between the nonreciprocity and the accompanying disorders in hopping amplitudes gives rise to energy-dependent localization transitions. Moreover, in the quasireciprocal Su-Schrieffer-Heeger models with staggered hopping $t_{jL}$ (or $t_{jR}$), topologically nontrivial phases are found in the real-spectra regimes characterized by nonzero winding numbers. Finally, we propose an experimental scheme to realize the quasireciprocal models in electrical circuits. Our findings shed new light on the subtle interplay among nonreciprocity, disorder, and topology.
\end{abstract}

%
%
%
\maketitle
%
%

\section{Introduction}

The past two decades have witnessed a fast growing interest in non-Hermitian (NH) systems for their intriguing properties and potential applications~\cite{Bender1998PRL,Bender2002PRL,Bender2007RPP,Moiseyev2011Book,Konotop2016RMP,Ganainy2018NatPhy,Ashida2020AiP,Bergholtz2021RMP}. NH terms in Hamiltonians may arise from the interaction with the environment in open systems~\cite{Rotter1991RPP,Rotter2009JPA}, the finite lifetime of quasiparticles~\cite{Fu2017arxiv,Shen2018PRL1,Yoshida2018PRB}, the complex refractive index~\cite{Musslimani2008PRL,Moiseyev2008PRL,Feng2017NatPho}, and the engineered Laplacian in electrical circuits~\cite{Schindler2011PRA,Luo2018arxiv,Lee2018ComPhy}. In contrast to Hermitian systems, the energy spectra of NH Hamiltonians are normally complex. One salient feature of the spectral theory of NH systems is the exceptional point, which is found in non-diagonalizable Hamiltonian matrices with varying parameters~\cite{Heiss2012JPAMT}. A more prominent aspect of NH Hamiltonians is the existence of real energy spectra~\cite{Bender2007RPP}. For instance, Bender et. al. showed that NH systems with $\mathcal{PT}$-symmetry can host real spectra~\cite{Bender1998PRL}. Mostafazadeha further proved that pseudo-Hermiticity is the necessary condition for the reality of spectrum~\cite{Mostafazadeh2002JMP,Mostafazadeh2010IJMMP}.

Recently, NH topological systems have been extensively studied both theoretically and experimentally~\cite{Bergholtz2021RMP,Rudner2009PRL,Esaki2011PRB,Bardyn2013NJP,Poshakinskiy2014PRL,Zeuner2015PRL,Malzard2015PRL,Aguado2016SciRep,Lee2016PRL,Molina2016PRL,Joglekar2016PRA,Zeng2016PRA,Weimann2017NatMat,Leykam2017PRL,Xu2017PRL,Menke2017PRB,Xiao2017NatPhy,Lieu2018PRB,Zyuzin2018PRB,Fan2018PRB,Alvarez2018PRB,Zhou2018Science,Yin2018PRA,Xiong2018JPC,Shen2018PRL2,Kunst2018PRL,Yao2018PRL1,Yao2018PRL2,Gong2018PRX,Kawabata2018PRB,Takata2018PRL,Chen2018PRB,Yi2018iScience,Hu2019PRB,Wang2019PRB,Rechtsman2018NatPho,Song2019PRB,Kunst2019PRB,Sato2019PRX,Zhou2019PRB,
Kawabata2019NatCom,Herviou2019PRA,Liu2019PRL,Edvardsson2019PRB,Luo2019PRL,Lee2019PRL,Yokomizo2019PRL,Song2019PRL,Zeng2020PRB1,Kawabata2020PRB,Zeng2020PRB2,Yang2020PRL,Okuma2020PRL,Zhang2020PRL,Borgnia2020PRL,Helbig2020NatPhy,Hofmann2020PRR,Xiao2020NatPhy,Weidemann2020Science,Chang2020PRR,Hu2021PRL,Chen2021SciPost}. The interplay between non-Hermiticity and topology induces numerous topological phenomena without Hermitian counterparts, e.g., the Weyl exceptional ring~\cite{Xu2017PRL}, the anomalous edge mode~\cite{Lee2016PRL,Liu2019PRL,Edvardsson2019PRB,Luo2019PRL}, and the point gap~\cite{Gong2018PRX}. One of the most interesting phenomena is the NH skin effect, in which the bulk states are localized at the boundaries by the nonreciprocal hopping, leading to the breakdown of the conventional bulk-boundary correspondence principle in topological systems~\cite{Yao2018PRL1}. In addition, nonreciprocity can also induce delocalization effect in the Anderson localization phase transition in quasiperiodic and disordered lattices~\cite{Hatano1996PRL,Hatano1997PRB,Amir2016PRE,Zeng2017PRA,Longhi2019PRL,Jiang2019PRB,Longhi2019PRB,Zeng2020PRR,Liu2020PRB,Liu2021PRB}.

So far, most studies have been focusing on NH systems with determined or uniform nonreciprocity. However, if the nonreciprocal hopping itself becomes aperiodic or disordered, what will happen to the systems' energy spectra, Anderson localization transition, and topological phases remain unexplored.  

In this paper, we introduce a one-dimensional (1D) quasireciprocal lattice model where the reciprocity is broken locally but recovered globally. The forward hopping amplitudes between the nearest neighboring lattice sites $\{ t+t_{jR} \}$ are chosen to be random permutations of the backward hopping $\{ t+t_{jL} \}$ or vice versa, so that $\sum_j t_{jR} = \sum_j t_{jL}$ always holds. For lattices with $\{ t_{jL} \}$ (or $\{t_{jR} \}$) being periodic, quasiperiodic, or disordered, the NH skin effect is absent due to the global cancellation of local nonreciprocity. We show that the Hamiltonian matrices are pseudo-Hermitian and the energy spectra are real as long as $\{ t_{jL} \}$ (or $\{t_{jR} \}$) are smaller than the threshold value. We also find that the eigenstates exhibit energy-dependent localization transitions as a result of the competition between the nonreciprocity and the accompanying disorders in hopping amplitudes. In addition, in the quasireciprocal Su-Schrieffer-Heeger models with staggered hopping $t_{jL}$ (or $t_{jR}$), the chiral symmetry is preserved and topological phases with zero-energy edge modes exist in the real-spectra regimes characterized by nonzero winding numbers. With stronger nonreciprocity, the nontrivial phases will be destroyed along with the pseudo-Hermiticity. Finally, we propose an experimental scheme to realize our model in electrical circuits. Our work unveils the subtle interplay among nonreciprocity, disorder, and topology.

The rest of the paper is organized as follows. In Sec.~\ref{sect2} we introduce the model Hamiltonian of the 1D quasireciprocal lattices. Then we discuss the real spectra and the pseudo-Hermiticity in the system in Sec.~\ref{sect3}. We further explore the Anderson localization phenomenon and topological phases in the quasireciprocal lattices in Sec.~\ref{sect4} and Sec.~\ref{sect5}, respectively. Finally, in Sec.~\ref{sect6} we propose an experimental scheme for realizing our model by employing electrical circuits. The last section (Sec.~\ref{sect7}) is dedicated to a brief summary.

\section{Model Hamiltonian}\label{sect2}
The 1D quasireciprocal lattice model is described by the following Hamiltonian
\begin{equation}\label{H}
	H = \sum_{j=1}^{N-1} (t+t_{jL}) c_j^\dagger c_{j+1} + (t+t_{jR}) c_{j+1}^\dagger c_j,
\end{equation}
where $c_j^\dagger$ ($c_j$) is the creation (annihilation) operator of spinless fermion at site $j$. ($t+t_{jL}$) and ($t+t_{jR}$) are respectively the backward and forward hopping amplitudes between two nearest neighboring sites. $t$ is a constant and will be taken as the energy unit throughout this paper. The values of $t_{jL}$ and $t_{jR}$ are chosen as follows. Firstly we set the backward hopping amplitudes as $\{t_{jL}\}=\{ t_{1L}, t_{2L},\cdots,t_{N-1,L} \}$ with $N$ being the number of lattice sites. Then the forward hopping amplitudes are chosen from the set $\{t_{jR}\}=\{ t_{1R}, t_{2R},\cdots,t_{N-1,R} \}=\mathcal{P} \{ t_{jL} \}$, which is a random permutation of $\{t_{jL}\}$. We can also set $\{t_{jR}\}$ first and take $\{t_{jL}\}$ as a random permutation of $\{t_{jR}\}$, but the conclusions are the same.

Since $t_{jR} \neq t_{jL}$, the forward and backward hopping between two neighboring sites can be different, which results in nonreciprocity. However, $\{t_{jR}\}$ is just a permutation of $\{t_{jL}\}$ and we always have $\sum_j t_{jR} = \sum_j t_{jL}$, the local nonreciprocity is canceled globally. To distinguish from the regular nonreciprocal lattices with uniform nonreciprocity, we call our model quasireciprocal lattices. The hopping amplitudes $\{t_{jL}\}$ in the model can be periodic, quasiperiodic, or randomly distributed. In the following, we will investigate the properties of quasireciprocal systems with $\{t_{jL}\}$ chosen in the following  way:
\begin{equation}\label{tL}
\eqalign{
Random:\quad t_{jL} \in (-\lambda,\lambda) \quad uniform \quad distribution; \cr
Quasiperiodic:\quad t_{jL} = \lambda \cos (2\pi \alpha j + \phi) \quad (\alpha=\frac{\sqrt{5}-1}{2}); \cr
Peirodic:\quad t_{jL} = \lambda \cos (2\pi \alpha j + \phi) \quad (\alpha=p/q). }
\end{equation}
Here $\lambda$ is a real number, $p$ and $q$ are co-prime integers, and $\phi$ is the phase of modulation in the periodic cases. 

\begin{figure}[t]
	\centering
	\includegraphics[width=4.0in]{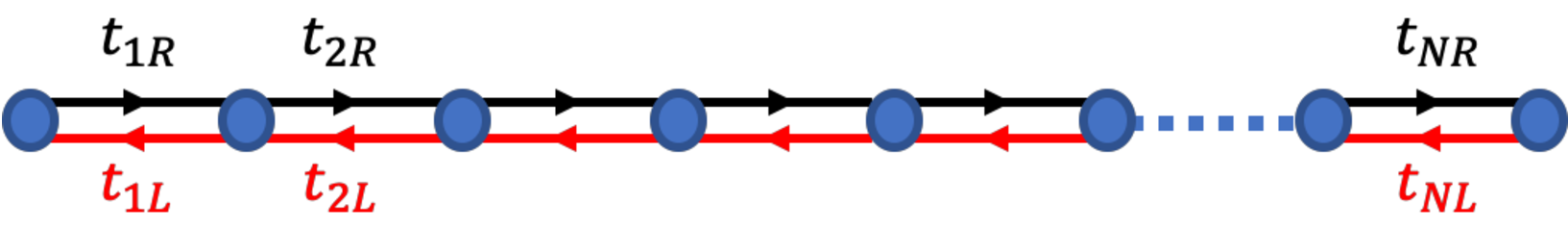}
	\caption{(Color online) Schematic illustration of the one-dimensional quasireciprocal lattices. The forward hopping amplitudes $\{ t_{jR} \}$ are chosen to be a random permutation of the backward hopping amplitudes $\{ t_{jL} \}$ or vice versa.}
	\label{fig1}
\end{figure}

\section{Real spectra and pseudo-Hermiticity}\label{sect3}
\begin{figure}[t]
	\centering
	\includegraphics[width=4.0in]{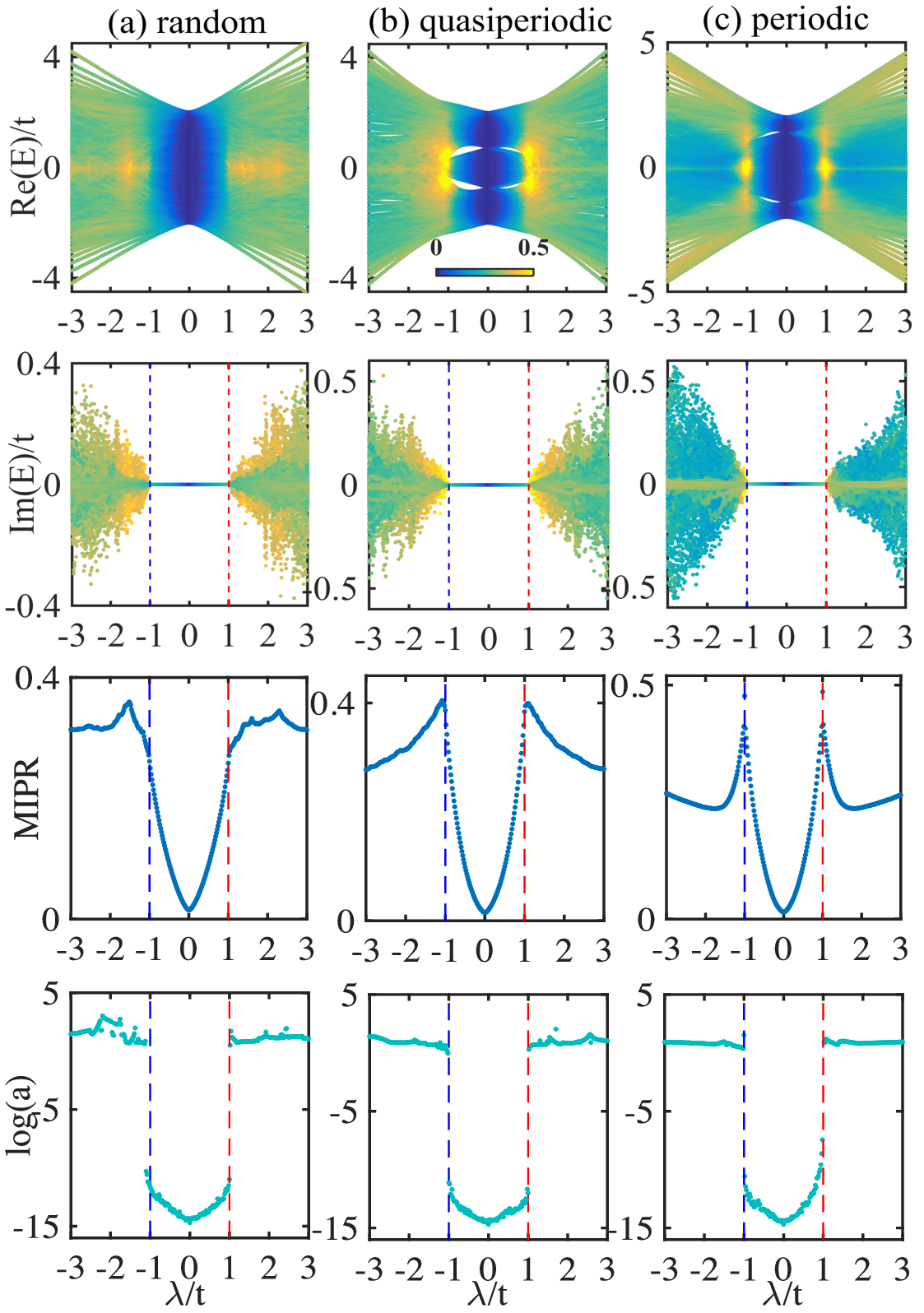}
	\caption{(Color online) Energy spectra and MIPR for the quasireciprocal lattices under open boundary conditions  with $t_{jL}$ being (a) uniformly distributed in $(-\lambda,\lambda)$, (b) quasiperiodic [$\alpha=(\sqrt{5}-1)/2, \phi=0$], and (c) periodic ($\alpha=1/4, \phi=0$). The first and second rows show the real and imaginary parts of the eigenenergies as a function of $\lambda$. The color bar indicates the IPR values of the eigenstates. The MIPR values of all the eigenstates for the system with different $\lambda$ are shown in the third row. The results are obtained by averaging over 100 samples with different permutations of $\{ t_{jL} \}$. The lattice size is $N=100$. The lowest row presents the logarithm of the maximum element in matrix $H^\dagger - \eta_R^{-1} H \eta_R$.}
	\label{fig2}
\end{figure}

We first check the energy spectra of the model shown in Eq.~(\ref{H}). Since the Hamiltonian matrix of the system is non-Hermitian, we may expect the spectrum to be complex. Interestingly, the eigenenergies can be real for certain parameter regimes. In the first and second rows of Fig.~\ref{fig2}, we present the real and imaginary parts of the eigenenergies for the lattices under open boundary conditions. The results are obtained by averaging over 100 samples where the forward hopping $\{ t_{jR} \}$ are different permutations of the backward hopping $\{ t_{jL} \}$. From the imaginary parts we can find that for all the three cases defined in Eq.~(\ref{tL}), the eigenenergies are purely real for $|\lambda|<t$, as indicated by the blue and red dashed lines in the second row. When $\lambda$ gets stronger than the critical value $\lambda_c = \pm t$, the spectra become complex. Next, we demonstrate that the real spectra of quasireciprocal lattices originate from the pseudo-Hermiticity of the Hamiltonian matrices. 

\begin{figure}[t]
	\centering
	\includegraphics[width=4.0in]{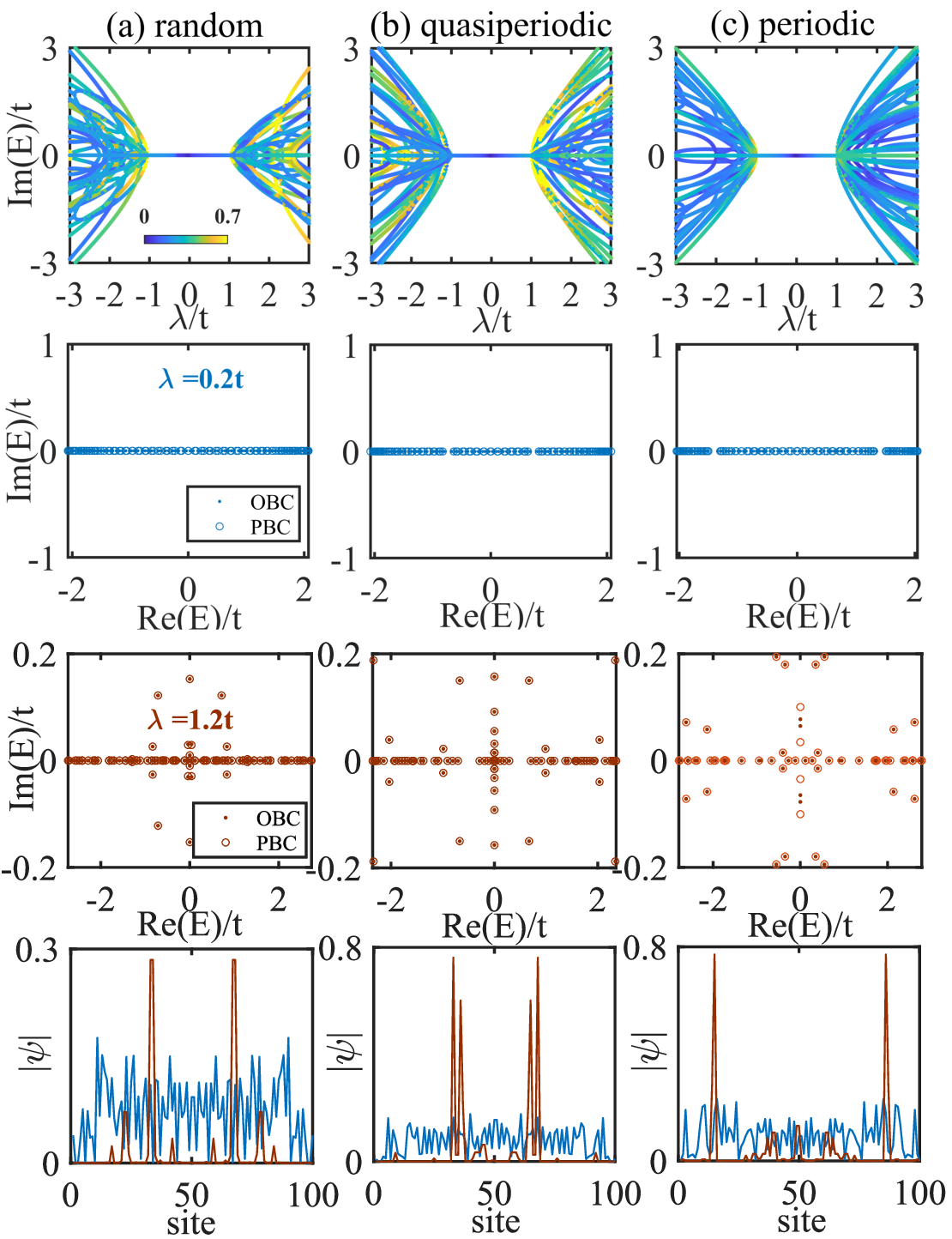}
	\caption{(Color online) Energy spectra and eigenstates of the quasireciprocal lattices calculated for only one sample. The first row shows the imaginary parts of energy spectra for the quasireciprocal lattices with $t_{jL}$ being (a) uniformly distributed in $(-\lambda,\lambda)$, (b) quasiperiodic [$\alpha=(\sqrt{5}-1)/2, \phi=0$], and (c) periodic ($\alpha=1/4, \phi=0$). The second and third rows show the OBC (solid dots) and PBC (empty circles) spectra of the systems with $\lambda=0.2t$ and $\lambda=1.2t$, respectively. The fourth row illustrates the space distribution of the eigenstate at $\lambda=0.2t$ (blue lines) and $\lambda=1.2t$ (brown lines).}
	\label{fig3}
\end{figure}

For a non-Hermitian Hamiltonian $H$, we have 
\begin{equation}
	H | \psi_{nR} \rangle = E_n | \psi_{nR} \rangle, \quad  H^\dagger | \psi_{nL} \rangle = E_n^* | \psi_{nL} \rangle,
\end{equation}
where $| \psi_{nR} \rangle$ and $| \psi_{nL} \rangle$ are the right and left eigenstates corresponding to the $n$th eigenenergy $E_n$ and its conjugate $E_n^*$, respectively. The eigenstates satisfy the biorthonormal relation $\langle \psi_{nL} | \psi_{mR} \rangle = \delta_{nm}$ and compromise a complete basis $\sum_n | \psi_{nR} \rangle \langle \psi_{nL} | = \sum_n | \psi_{nL} \rangle \langle \psi_{nR} | = \bold{1}$ after normalization. To prove that the Hamiltonian matrix is pseudo-Hermitian, we construct two $\eta$ matrices as 
\begin{equation}
	\eta_R = \sum_n | \psi_{nR} \rangle \langle \psi_{nR} |, \quad \eta_L = \sum_n | \psi_{nL} \rangle \langle \psi_{nL} |.
\end{equation} 
Then we can check whether the following relations are satisfied
\begin{equation}
	H^\dagger = \eta_R^{-1} H \eta_R, \qquad H = \eta_L^{-1} H^\dagger \eta_L.
\end{equation}
For the biorthonormal eigenstates $| \psi_{nR} \rangle$ and $| \psi_{nL} \rangle$, we have $\eta_L = \eta_R^{-1}$. Since $\{ t_{jR} \}$ are random permutations of $\{ t_{jL} \}$, the analytic proof of pseudo-Hermiticity is quite difficult. Nevertheless, we can numerically determine whether the matrix $H^\dagger-\eta_R^{-1} H \eta_R$ (or $H-\eta_L^{-1} H^\dagger \eta_L$) is a null matrix or not. Take the $\eta_R$ case as an example, we define a new variable 
\begin{equation}
	a = max\{ H^\dagger-\eta_R^{-1} H \eta_R \},
\end{equation}
such that $a$ is the maximum element in the matrix. In the last row of Fig.~\ref{fig2}, we show the logarithm of $a$ as a function of $\lambda$ for the three kinds of quasireciprocal lattices. When $|\lambda|<t$, $a$ is almost zero (around the order of $10^{-10}-10^{-15}$), indicating that $H^\dagger = \eta_R^{-1} H \eta_R$ and the Hamiltonian is pseudo-Hermitian. At the critical value $\lambda_c=\pm t$, $a$ jumps sharply to a much larger value of the order $O(1)$, implying the breaking of pseudo-Hermiticity. Then the spectrum becomes complex for stronger $\lambda$. Similar results are also obtained for $H-\eta_L^{-1} H^\dagger \eta_L$. Thus we conclude that the real spectra in our model arise from the pseudo-Hermiticity in the Hamiltonian. Notice that though our results in Fig.~\ref{fig2} are obtained by averaging different samples, the conclusion still holds for a single sample. In Fig.~\ref{fig3}, we present the imaginary parts of the spectra of one sample for the different kinds of quasireciprocal lattices. Clearly, the spectra are entirely real when $|\lambda|<t$, but becomes complex when $|\lambda|>t$.

It is known that the emergence of non-Hermitian skin effect (NHSE) in the nonreciprocal systems is closely connected to the point gap in the energy spectra under periodic boundary conditions (PBCs)~\cite{Okuma2020PRL,Zhang2020PRL,Borgnia2020PRL}. As to the quasireciprocal lattices we discuss here, we find that there is no skin effect in the eigenstates. From the IPR values of eigenstates shown in Fig.~\ref{fig2}, we can see that there are extended states when $\lambda$ is small. Since the spectra and IPR are obtained under OBCs, it means that these states will not be shifted to the boundaries. The second and third rows of Fig.~\ref{fig3} are the OBC and PBC spectra at $\lambda=0.2t$ and $\lambda=1.2t$. We find that the spectra under OBCs and PBCs are almost the same, and there is no loop structures and point gaps in the PBC spectra. In the lowest row of Fig.~\ref{fig3}, we show the space distribution of the eigenstates under OBCs, we can see that the eigenstates are extended for small $\lambda$ but become localized for large $\lambda$. The NHSE is absent in system, which is consistent with the results that no skin effect exist in the system under OBCs. This is understandable since the nonreciprocity in the forward and backward hopping amplitudes are always canceled with each other.

\section{Anderson localization}\label{sect4}
Another interesting phenomenon we can observe in our model is the Anderson localization. In the quasireciprocal lattices, the forward hopping amplitudes $\{ t_{jR} \}$ are random permutations of the backward hopping amplitudes $\{ t_{jL} \}$, such disorders will localize the eigenstates. To characterize the localization properties, we define the inverse participation ratio (IPR) for each eigenstate as $IPR(E_n)=\sum_j^{N} |\psi_{nR,j}|^4/[\sum_j |\psi_{nR,j}|^2]^2$, where $\psi_{nR,j}$ represents the $j$th component of the right eigenvector $| \psi_{nR} \rangle$ with energy $E_n$. The IPR values are of the order $O(1/L)$ for extended states but become of order $O(1)$ for localized states. In Fig.~\ref{fig2}, the IPR values of eigenstates are indicated by the color bar. We find that for small $\lambda$ values, there are extended states in the system. In the lowest row of Fig.~\ref{fig3}, we present the distribution of the eigenstates at $\lambda=0.2t$ and $1.2t$, where the extened state and localized state are found. As $\lambda$ increases, more and more states become localized. The localization is energy-dependent, where the states near the band edges are easier to be localized than those near the band center. In contrast to disorders, it is known that nonreciprocal hopping can induce delocalization effect~\cite{Gong2018PRX,Jiang2019PRB}. Here in our model, the presence of nonreciprocity is always accompanied by random disorders in the hopping amplitudes, the competition between these two factors thus leads to the extended-to-localized-state transitions. As $\lambda$ gets stronger, the effect of disorder overtakes that of nonreciprocity, and all the eigenstates become localized.

To better illustrate the localization properties in the quasireciprocal lattices, we further calculate the mean inverse participation ratio $MIPR=\sum_{n=1}^N IPR(E_n)/N$ for each $\lambda$ value, which are presented in the third row in Fig.~\ref{fig2}. As $\lambda$ grows, the MIPR gradually ramps up instead of sharply jumping to a large value, implying that not all the states are localized at the same time. For the MIPR values of quasiperiodic and periodic cases shown in Fig.~\ref{fig2}(b) and \ref{fig2}(c), two peaks emerge at $\lambda=\pm t$. So the states are more localized there, corresponding to the bright yellow regions present in the real parts of spectra. The reason behind this is that when $\lambda \approx \pm t$ in the (quasi)periodic cases, the hopping amplitudes between certain sites are very close to $0$, resulting in states localized on those sites. Such phenomenon is less obvious in the random case since the hopping amplitudes are randomly chosen. It is interesting to see that the critical values for the pseudo-Hermiticity are also connected to the localization properties. 

\section{Topological phase}\label{sect5}
\begin{figure}[t]
	\centering
	\includegraphics[width=4.0in]{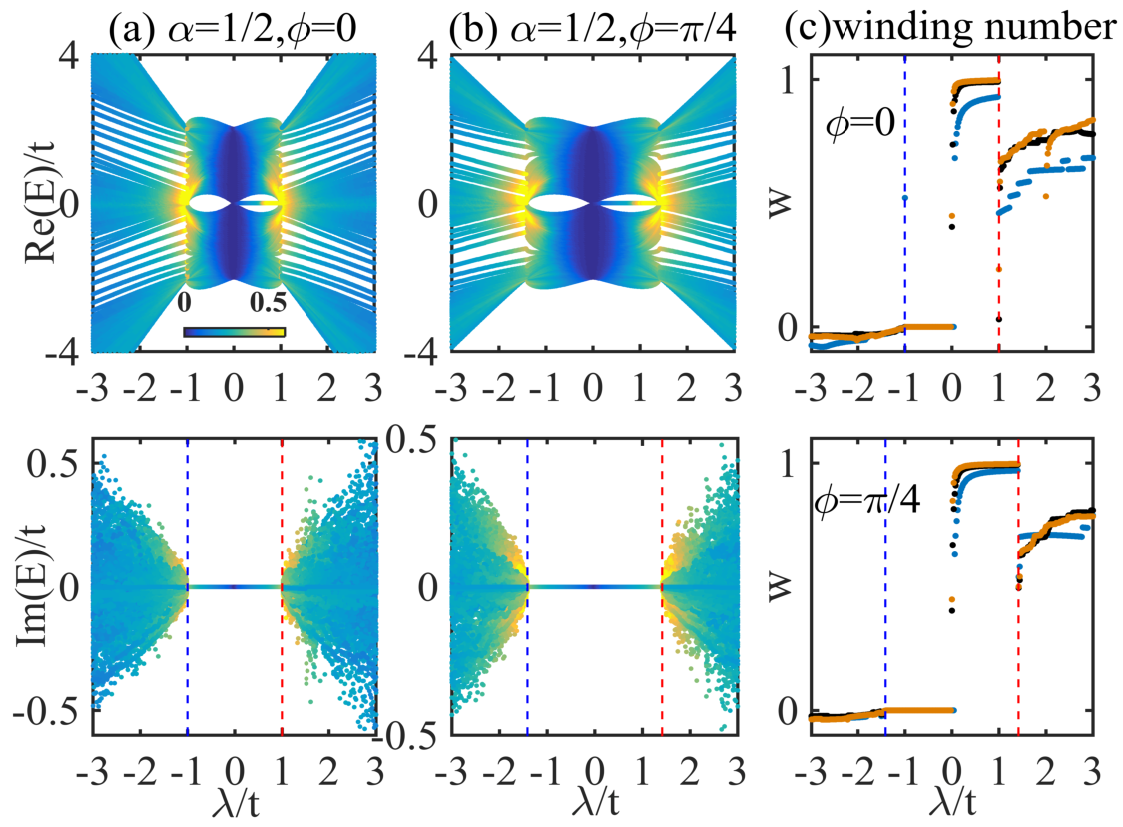}
	\caption{(Color online) Energy spectra for the quasireciprocal SSH model where $t_{jL}$ is periodic with (a) $\alpha=1/2, \phi=0$ and (b) $\alpha=1/2, \phi=\pi/4$. The color bar indicates the IPR values of eigenstates. The lattice size is $N=100$ and the results are averaged over 100 samples. (c) shows the winding numbers as a function of $\lambda$ for these two cases, respectively. The colored dots represent the lattices with different sizes: $N=100$ (blue), $N=400$ (black), and $N=1000$ (yellow).}
	\label{fig4}
\end{figure}

Now we turn to the topological phases in the lattices with periodic backward (or forward) hopping amplitudes $t_{jL}$ (or $t_{jR}$). Without loss of generalities, we set $t_{jL}$ to be periodic. First we check the case with $\alpha=1/2$, then the backward hopping amplitudes are
\begin{equation}
t_{jL}=\left\{ \eqalign{
t-\lambda \cos(\phi) \quad j \quad odd; \cr
t+\lambda \cos(\phi) \quad j \quad even.}
\right.
\end{equation}
Thus we obtain staggered hopping terms in the 1D lattice, similar to the Su-Sherieffer-Heeger (SSH) model. The difference is that here the forward hopping $\{ t_{jR} \}$ is a random permutation of $\{ t_{jL} \}$. In the normal SSH model, it is well known that the system is topologically nontrivial when $\lambda > 0$. For the quasireciprocal SSH model studied here, since the Hamiltonian is non-Hermitian and there are disorders in the system, the situation becomes quite different. In Fig.~\ref{fig4}(a), we show the energy spectrum for the system with $\alpha=1/2$ and $\phi=0$, which is obtained by averaging over 100 samples. From the IPR values of the eigenstates, we know there are extended states when $\lambda$ is relatively small, similar to the cases we discussed in Fig.~\ref{fig2}. It is interesting to find that when $|\lambda|<t$, there is an energy gap in the spectrum. The eigenenergies in this regime are real as can be seen from the vanishing parts. Furthermore, zero-energy edge modes exist in the regime $0<\lambda<t$. As $\lambda$ becomes stronger, the energy gap will be closed and the topological phase is destroyed. If we set $\phi=\pi/4$ in the model, then the parameter regime with real spectra will change to $|\lambda|<\sqrt{2}t$, and the nontrivial regime also expands to $0<\lambda<\sqrt{2}t$, as shown in Fig.~\ref{fig4}(b). In general, the regime with real spectrum for the quasireciprocal lattices with periodic $t_{jL}$ (or $t_{jR}$) is $|\lambda|<t/\cos(\phi)$. The topologically nontrivial regime for the quasireciprocal SSH model is $0<\lambda<t/\cos(\phi)$. So, the existence of the topological phase is accompanied by the pseudo-Hermiticity in our quasireciprocal model.

To characterize the topological phase in the quasireciprocal SSH models, we can use winding numbers. Since the lattices are not periodic, we have to calculate them numerically in real space~\cite{Song2019PRL,Song2014PRB,Shem2014PRL,Lin2021PRB}. We find that even though the system is disordered, the chiral symmetry is still preserved in the quasireciprocal lattices with $q$ being even, i.e., we have $S^{-1}HS = -H$ with $S=diag(1,-1,1,-1,\cdots)$. For non-Hermitian systems, we can construct the $Q$ matrix as~\cite{Song2019PRL}
\begin{equation}
	Q = \sum_{n=1}^{N} ( |\psi_{nR}\rangle \langle \psi_{nL}| - |\tilde{\psi}_{nR}\rangle \langle \tilde{\psi}_{nL}| ),
\end{equation}
where $| \tilde{\psi}_{n,L/R} \rangle = S |\psi_{n,L/R}\rangle$. The lattice size is set to be $N=qN'$ with $N'$ being the number of quasicell including the $q$ nearest sites. Then the winding number $W$ in the real space is defined as
\begin{equation}
	W = \frac{1}{2N} Tr (SQ[Q,X]).
\end{equation}
Here, $X$ is the coordinate operator, i.e., $X=diag(1,2,3,\cdots,N') \otimes I_{q \times q}$ with $I_{q \times q}$ being a unit matrix of dimension $q \times q$. In Fig.~\ref{fig4}(c), we present the numerical results of the winding numbers for the quasireciprocal SSH models with $\phi=0$ and $\pi/4$, respectively. We see that in the regime with real eigenenergies, the winding number is quantized to integers as the system size increases. We have $W=0$ for $-t/\cos(\phi)<\lambda<0$ and $W=1$ for $0<\lambda<t/\cos(\phi)$, consistent with the regime where zero-energy edge modes present. While if $|\lambda|>t/\cos (\phi)$, the winding numbers are not integers. Thus we show that in the quasireciprocal SSH models, we can obtain topologically nontrivial phases with zero-energy edge modes.

For other quasireciprocal lattices with $q>2$, we find that there are no topological phases after averaging over different samples. For instance, the spectrum for the lattice with $\alpha = 1/4$ is shown in Fig.~\ref{fig2}(c), where the energy gaps are smeared out and no midgap modes exist after the averaging. We argue that for $q>2$, the disorder in the hopping amplitudes within the quasicell will wipe out the topological phases. While for the lattice with $q=2$, there are only two values for the hopping amplitudes and the disorder is relatively weak. The nontrivial phase can exist in a wide parameter regime before being destroyed by stronger local nonreciprocity, which essentially enhances the disorder in the system.

\begin{figure}[t]
	\centering
	\includegraphics[width=4.0in]{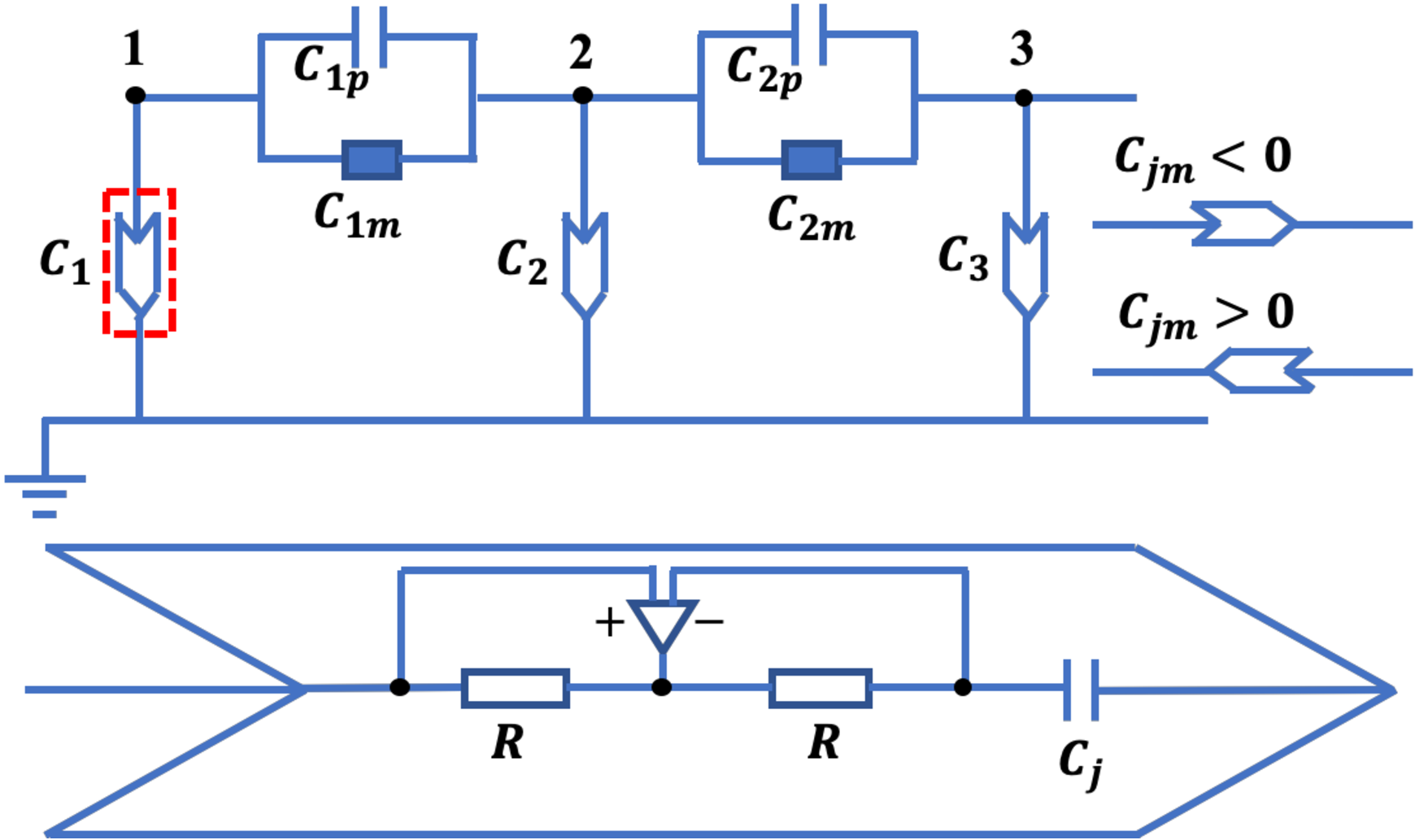}
	\caption{(Color online) Upper panel: Electrical circuit for realizing the quasireciprocal lattice models. $C_{jp}$ is a capacitor and $C_{jm}$ is an INIC, which are combined to simulate the nonreciprocal hopping. The Lower panel shows the structure of the INIC contained in the red rectangle, which consists of two resistors, one capacitor, and one operational amplifier. When the current runs from left to right, the capacitance of the INIC is $-C_j$; if the current reverses, the capacitance will be $C_j$.}
	\label{fig5}
\end{figure} 

\section{Experimental realization}\label{sect6}
Recently, electrical circuit has become a powerful and versatile platform to simulate various tight-binding lattice models~\cite{Luo2018arxiv,Lee2018ComPhy,Helbig2020NatPhy,Hofmann2020PRR,Hofmann2019PRL,Dong2021PRR}. Here we propose an experimental scheme to realize the quasireciprocal lattices by using electrical circuits, as shown in Fig.~\ref{fig5}. The nonreciprocal hopping between the nearest lattice sites are simulated by combining a normal capacitor $C_{jp}$ and a negative impedance converter with current inversion (INIC) $C_{jm}$~\cite{Chen2009Book}. The INIC consists of a capacitor, two resistors, and one operational amplifier [see the structure shown in the lower panel in Fig.~\ref{fig4}]. When the current runs through the INIC from the left to the right, the capacitance of the INIC is $-C_j$. If the current runs in the opposite direction, i.e., from right to the left, the capacitance will be $C_j$. Suppose that the input current and electrical potential at node $j$ are $I_j$ and $V_j$, respectively. Then according to Kirchhoff's law, we have 
\begin{equation}
	I_j = \sum_i Y_{ji} (V_j - V_i) + X_j V_j, 
\end{equation}
where $i$ denotes all the nodes linked to node $j$ with conductance $Y_{ij}$. $X_j$ is the conductance of node $j$. Considering all the nodes, the relation between the currents and voltages can be expressed in a compact matrix form as 
\begin{equation}
	\bold{I}=\bold{JV},
\end{equation}
with $\bold{J}$ being the Laplacian of the circuit, which corresponds to the model Hamiltonian matrix. 

In the circuit, the neighboring nodes are connected by a normal capacitor $C_{jp}$ and an INIC $C_{jm}$. In addition, each node is grounded through another INIC $C_j$, which is used to eliminate the diagonal terms since we only have off-diagonal terms in the Hamiltonian. 
Thus we set
\begin{equation}
	\eqalign{
	C_1 = C_{1p}+C_{1m}, \qquad C_L = C_{L-1,p}-C_{L-1,m}, \cr
	C_j=C_{j-1,p}-C_{j-1,m}+C_{j,p}+C_{j,m} \quad for \quad 1<j<L. }
\end{equation}
Then the circuit Laplacian reduces to
\begin{equation}
	\boldsymbol{J}= i \omega
	\left(
	\begin{array}{ccccc}
		0 & -(C_{1p}+C_{1m}) & 0 & \cdots & 0 \\
		-(C_{1p}-C_{1m}) & 0 & -(C_{2p}+C_{2m}) & \cdots & 0 \\
		0 & -(C_{2p}-C_{2m}) & 0 & \cdots & 0 \\
		\vdots & \vdots & \vdots & \ddots & \vdots \\
		\cdots & \cdots & \cdots & \cdots & 0
	\end{array}
	\right).
\end{equation}

To model the non-Hermitian Hamiltonian in the main text, we further set
\begin{equation}
	C_{jp} = t + (t_{jR} + t_{jL})/2, \qquad C_{jm} = (t_{jR} - t_{jL})/2,
\end{equation}
where $t$ is constant hopping amplitude, $t_{jR}$ and $t_{jL}$ are the forward and backward hopping amplitudes, respectively. Then the forward hopping ($C_{jp}+C_{jm}=t+t_{jR}$) and backward hopping ($C_{jp}-C_{jm}=t+t_{jL}$) are asymmetric. Hence the Hamiltonian for the 1D quasireciprocal lattice is achieved. The energy spectrum of the system can be obtained from the admittance spectrum of the circuit, and the distribution of edge states can be detected by measuring the voltage at each node.

\section{Summary}\label{sect7}
In summary, we have introduced a new 1D non-Hermitian model called quasireciprocal system. We show that the energy spectra are real when the pseudo-Hermiticity is preserved in the model Hamiltonians. The competition between disorder and nonreciprocity gives rise to the energy-dependent localization transitions in eigenstates. We also find that topologically nontrivial phases can exist in the quasireciprocal SSH models. Finally, we propose an experimental scheme to realize the quasireciprocal model by employing electrical circuits. Our model reveals the exotic properties of non-Hermitian systems and the subtle interplay among nonreciprocity, disorder, and topology.

\section*{Acknowledgments}
This work is supported by the Open Research Fund Program of the State Key Laboratory of Low-Dimensional Quantum Physics. R. L\"u is supported by NSFC under Grants No. 11874234 and the National Key Research and Development Program of China (2018YFA0306504).


\end{document}